\documentclass{aa}
\usepackage{graphicx}
\usepackage{amsmath}   
\usepackage{amssymb}
\usepackage{times}

\usepackage{color}

\def\eq#1{{Eq.~(\ref{#1})}}

\begin{document}

\title{Unravelling the formation of the first supermassive black holes with the SKA pulsar timing array}
\titlerunning{SMBHs with SKA PTA}

 \author{Hamsa Padmanabhan
          \inst{1}
          \and
          Abraham Loeb\inst{2}
          }

   \institute{D\'epartement de Physique Th\'eorique, Universit\'e de Gen\`eve \\
24 quai Ernest-Ansermet, CH 1211 Gen\`eve 4, Switzerland\\
              \email{hamsa.padmanabhan@unige.ch}
         \and
            Astronomy department, Harvard University \\
60 Garden Street, Cambridge, MA 02138, USA \\
             \email{aloeb@cfa.harvard.edu}            
             }

   \date{}


\abstract
 {Galaxy mergers at high redshifts trigger the activity of their central supermassive black holes, eventually also leading to their coalescence  -- and a potential source of low-frequency gravitational waves detectable  by  the SKA Pulsar Timing Array (PTA). Two key parameters related to the fuelling of black holes are the Eddington ratio of quasar accretion $\eta_{\rm Edd}$, and the radiative efficiency of the accretion process, $\epsilon$ (which affects the so-called active lifetime of the quasar, $t_{\rm QSO}$). We forecast the regime of detectability of gravitational wave events with SKA PTA, finding the associated binaries to have orbital periods on the order of weeks to years, observable through relativistic Doppler velocity boosting and/or optical variability of their light curves.  Combining the SKA regime of detectability with the latest observational constraints on  high-redshift black hole mass and luminosity functions, and theoretically motivated prescriptions for the merger rates of dark matter haloes, we forecast the number of active counterparts of SKA PTA events expected as a function of primary black hole mass at $z \gtrsim 6$. We find that the quasar counterpart of the most massive black holes will be {\it uniquely localizable} within the SKA PTA error ellipse at  $z \gtrsim 6$. We also forecast the number of expected counterparts as a function of the quasars' Eddington ratio and active lifetime. Our results show that SKA PTA detections can place robust constraints on the seeding and growth mechanisms of the first supermassive black holes.
 }


\maketitle

\section{Introduction}
\begin{figure}
\begin{center}
\includegraphics[width =\columnwidth]{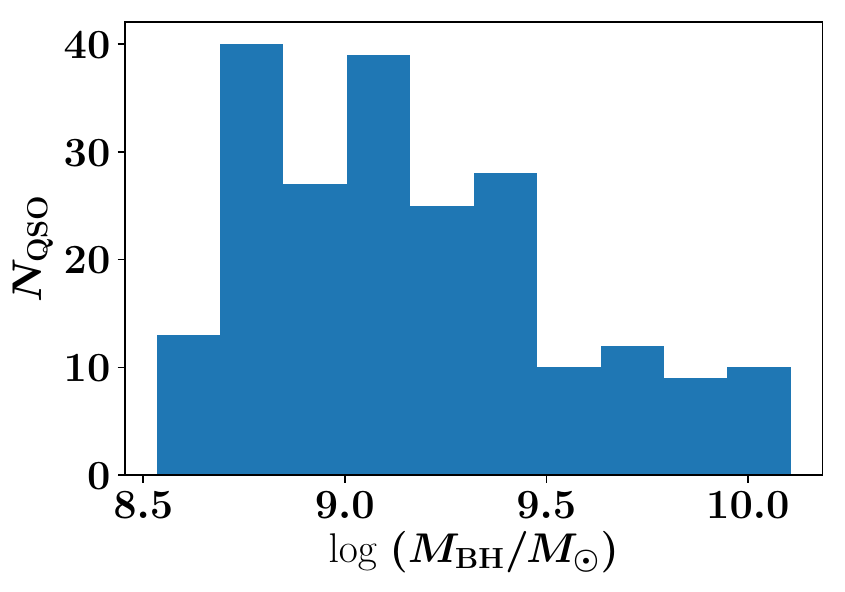}
\end{center}
\caption{Distribution of estimated black hole masses in observed $z > 6$ quasars. Masses are derived from bolometric luminosities, where available, assuming Eddington ratios from \citet{nobuta2012}. These are also consistent with mass estimates derived independently in the literature.}
\label{fig:highzmbh}
\end{figure}

 The seeding and growth of the first supermassive black holes in the Universe is an outstanding question in contemporary astrophysics \citep{rees1984, turner1991, barkana2001}.  We currently  have an observational  sample of 213 quasars\footnote{While the term `quasar' has traditionally been used to describe radio-loud quasi-stellar objects, we use the terms `QSOs' and `quasars' interchangeably in what follows, irrespective of their radio properties.}  at $z > 6$ from current surveys \citep[for the latest compilation and a `live' catalog, see, e.g.,][]{flesch2021}. The masses of the supermassive black holes associated with the observed quasars follow a broad distribution (Fig. \ref{fig:highzmbh}) between $10^{8} M_{\odot} < M_{\rm BH} < 10^{10} M_{\odot}$, with most observed to be in a rapidly accreting phase having Eddington ratios close to unity \citep{fan2006,  wu2015, kim2019,millon2022,  venemans2016, venemans2017, Willott_2015, 
trakhtenbrot2017, banados2018, feruglio2018, mortlock2011, derosa2014}.  For a reasonable range of seed values, the maximum mass of a black hole at any redshift is predicted to be $10^{10} M_{\odot}$ \citep{haiman2001, natarajan2009, inayoshi2016, king2016},  consistently with that of the most massive black hole at $z > 6$ discovered so far \citep[$\sim 1.2 \times 10^{10} M_{\odot}$;][]{wu2015}. 
When combined with the inferred lifetimes of quasars \citep{eilers2021, khrykin2021, eilers2020, volonteri2010, volonteri2012},  this leads to the question of how such massive systems grew over a relatively short period in the Universe's history \citep[for a review,  see][]{inayoshi2020}.

In addition to the black hole mass, two key physical  parameters that determine the black hole seeding and growth mechanisms are the bolometric luminosity $L_{\rm bol}$ of the associated quasar (usually expressed in units of the Eddington luminosity, $L_{\rm Edd} = 1.3 \times 10^{38} (M_{\rm BH}/M_{\odot}) L_{\odot}$, as $\eta_{\rm Edd} = L_{\rm bol}/L_{\rm Edd}$) and the radiative efficiency of accretion of material on to the black hole, $\epsilon$, which is believed to power nuclear activity and is  related \citep{pacucci2022}   to the active lifetime of the quasar, $t_{\rm QSO}$.  The combination of $\eta_{\rm Edd}$ and  $\epsilon$ is constrained by the Salpeter time-scale \citep{salpeter1964}
 for black hole growth, expressible as $t_{\rm S} = 0.45 \left({\epsilon}/{1-\epsilon}\right) \left({L_{\rm bol}}/{L_{\rm Edd}}\right)^{-1} \ {\rm Gyr}$. The mass of the black hole grows exponentially  according to $M_{\rm BH} = M_{\rm seed} \exp(t_{\rm QSO}/t_S)$ --
starting from a seed mass $M_{\rm seed}$, which could have different possible origins, from remnants of Pop III stars \citep{Madau_2001}, primordial black holes from the early Universe \citep{carr1974},  or `heavy' seeds formed from direct collapse of gas clouds \citep{haehnelt1993, bromm2003, lodato2006, lodato2007, agarwal2016, pacucci2017,natarajan2017,  bonoli2014}.

In the hierarchical structure formation scenario, galaxy mergers are expected to be frequent at $z \gtrsim 6$, leading to the inevitable formation of binary supermassive black holes (SMBHs). There is, however, considerable uncertainty in our understanding of the evolution of such systems \citep[e.g.,][]{begelman1980, armitage2002, milosavljevic2003, kulkarni2012}, especially their lifetimes and coalescence timescales. The latter of these is important in view of observational signatures such as gravitational wave emission which is expected to accompany the final stages of coalescence \citep{peters1964, loeb2010}.

Pulsar timing arrays aim to detect a gravitational wave background  at low frequencies by monitoring the spatially correlated fluctuations induced by the waves on the times of arrival (ToA) of radio pulses from millisecond pulsars. A key astrophysical source of such a  stochastic background is that generated by a cosmic population of SMBH binaries \citep[e.g.,][]{sesana2004, burke2019, maiorano2021, rajagopal1995, rosaldo2016, wen2011, bonetti2018, feng2020}, which emit gravitational waves with frequencies of the order of a few nHz once the separation reaches sub-pc scales.\footnote{More exotic sources include cosmic strings \citep{siemens2007, blanco2018, damour2001}, phase transitions \citep{caprini2010,kobakhidze2017} or a primordial GW background in the early Universe \citep{Grishchuk_2005, lasky2016}.}

The Square Kilometre Array (SKA) will revolutionize gravitational wave astronomy with pulsar timing \citep{stappers2018, kramer2004, janssen2015}, thanks to its much larger collecting area and improved sensitivity to frequencies of the order of the reciprocal of the observing time, i.e. nHz for observing timescales of a few years.   A Pulsar Timing Array (PTA) with the SKA is predicted to discover more than 14000 canonical and 6000 millisecond pulsars within 5 to 8 years of operation \citep{smits2009, taylor2016, CORDES_2004, rosado2015}, and  has the potential to bring down the precision on timing the residuals (computed from the phase difference between the observed ToA and the predicted ToA based on model parameters) to 1-10 ns \citep{liu2011, Spallicci_2013}.  Sensitive to frequencies between 1 nHz - 100 nHz,  SKA PTA will significantly boost \citep{kramer2004, janssen2015} the existing efforts from the International PTA collaboration \citep[IPTA;][]{verbiest2016} to detect low-frequency gravitational waves, complementing the regime probed by the LIGO, {Advanced Virgo and KAGRA} (Hz) and LISA (mHz) facilities.

In this paper, we outline the prospects for detecting  SMBH mergers at redshifts $z \gtrsim 6$ from their gravitational wave signal measured by the SKA PTA.  In so doing, we provide fully data-driven expressions for the proportion of active quasars among black holes at high redshifts, as well as the fraction of binary SMBHs.  The SMBH binaries' contribution to the background, expected to be the brightest in the PTA band, can be robustly isolated within a few years of integration \citep{pol2021, kaiser2022}. We find that SKA PTA should be sensitive to SMBHs with primary black hole masses of $M_{\rm BH} \gtrsim 10^9 M_{\odot}$, mass ratios greater {than $q_{\rm min} = 0.25$ (for $M_{\rm BH} \sim 10^9 M_{\odot}$) and $q_{\rm min} = 0.005$ (for $M_{\rm BH} \sim 10^{10} M_{\odot}$)} and separations $a = 0.5 - 50$ mpc,   fairly independently of redshift. Assuming that the black hole mergers proceed rapidly to coalescence, we forecast the expected number of detections and their electromagnetic counterparts [both quasars \citep{kocsis2006, Casey_Clyde_2022} and galaxies \citep{haiman2009}] detectable by current and  future facilities. We explore the implications of our findings for placing constraints on the parameters $\eta_{\rm Edd}$ and $t_{\rm QSO}$ which are related to the growth mechanisms of black holes. In so doing, we provide fully data-driven expressions for the proportion of active quasars among black holes at high redshifts, as well as the fraction of binary SMBHs as a function of observationally determined parameters. 

The paper is organized as follows. In Sec. \ref{sec:formalism}, we develop the formalism connecting the observed strain and frequency of SKA PTA measurements to the properties of high redshift SMBHs. Using theoretically motivated prescriptions for the merger rates of dark matter haloes with empirically derived constraints on their black hole masses, we forecast the number of binary SMBHs detectable by  SKA PTA via their gravitational wave emission. In  Sec. \ref{sec:counterparts}, we outline prospects for detecting the electromagnetic counterparts of these sources, both galaxies (Sec. \ref{sec:galaxies}) and QSOs (Sec. \ref{sec:qsos}), including predictions for the number of active QSO counterparts expected as a function of black hole properties at $z > 6$. We summarize our findings in a brief concluding section (Sec. \ref{sec:discussion}).

\section{Gravitational wave constraints on supermassive black holes}
\label{sec:formalism}

\begin{figure}
\begin{center}
\includegraphics[width =\columnwidth]{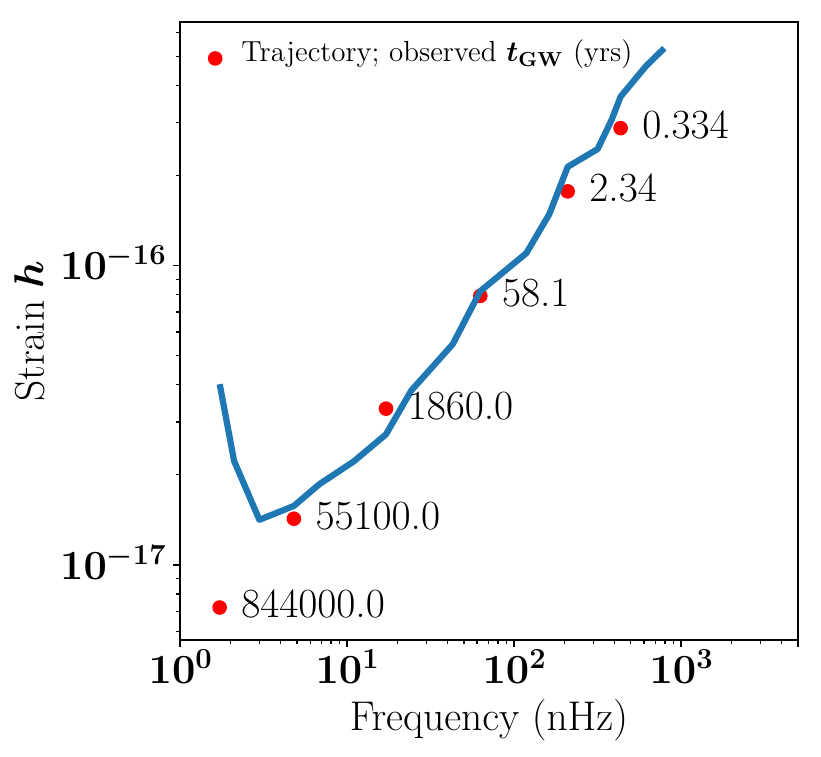}
\end{center}
\caption{Strain vs. observed frequency of SKA PTA \citep[blue curve, from][]{garcia2021}. The red dots mark example points on the trajectory of a binary system at $z \sim 6.8$ with masses $M_1 = 10^{10} M_{\odot}$ and $M_2 \approx 3 \times 10^7 M_{\odot}$ as a function of the gravitational wave decay timescale, $t_{\rm GW}$ measured in units of years, in the observer's frame of reference.}
\label{fig:hvsfska}
\end{figure}

The expected sensitivity of SKA PTA, as a function of the observed frequency, is shown by the blue curve in Fig. \ref{fig:hvsfska}, derived from the estimates of \citet{garcia2021} { which are based upon a nominal sensitivity level of $10^{-16} - 10^{-17}$ at the  reference frequency of 1 yr$^{-1}$.  This curve assumes 50 pulsars timed every two weeks for 20 years with a root-mean-square error in each timing residual of $30\;{\rm ns}$ from \citet{moore2015}. } Typically, the three observables, strain $h$, frequency $f_{\rm obs}$ and the gravitational wave decay timescale $t_{\rm gw, obs}$ may be measurable in the PTA survey. For an SMBH binary of masses $M_1$ and $M_2$ (assumed to be less than or equal to $M_1$) with a separation of $a$, and located at redshift $z$,  the strain and observed frequency are expressible as:
\begin{eqnarray}
h &=& \frac{2 (G \mathcal{M}_{\rm obs})^{5/3} (\pi f_{\rm obs})^{2/3}}{c^4 d_L} \nonumber \\
f_{\rm obs} &=& \frac{1}{\pi} \sqrt{\frac{G}{a^3}}\frac{(M_1 + M_2)^{1/2}}{(1+z)}
\label{freqstrain}
\end{eqnarray}
 In the above equations, $\mathcal{M}_{\rm obs}$ denotes the  observed chirp mass, defined by $\mathcal{M}_{\rm obs} = (M_1 M_2)^{3/5} (1+z)/(M_1 + M_2)^{1/5}$ and $d_L$ is the luminosity distance to redshift $z$.

\begin{figure}
\begin{center}
\includegraphics[width =\columnwidth]{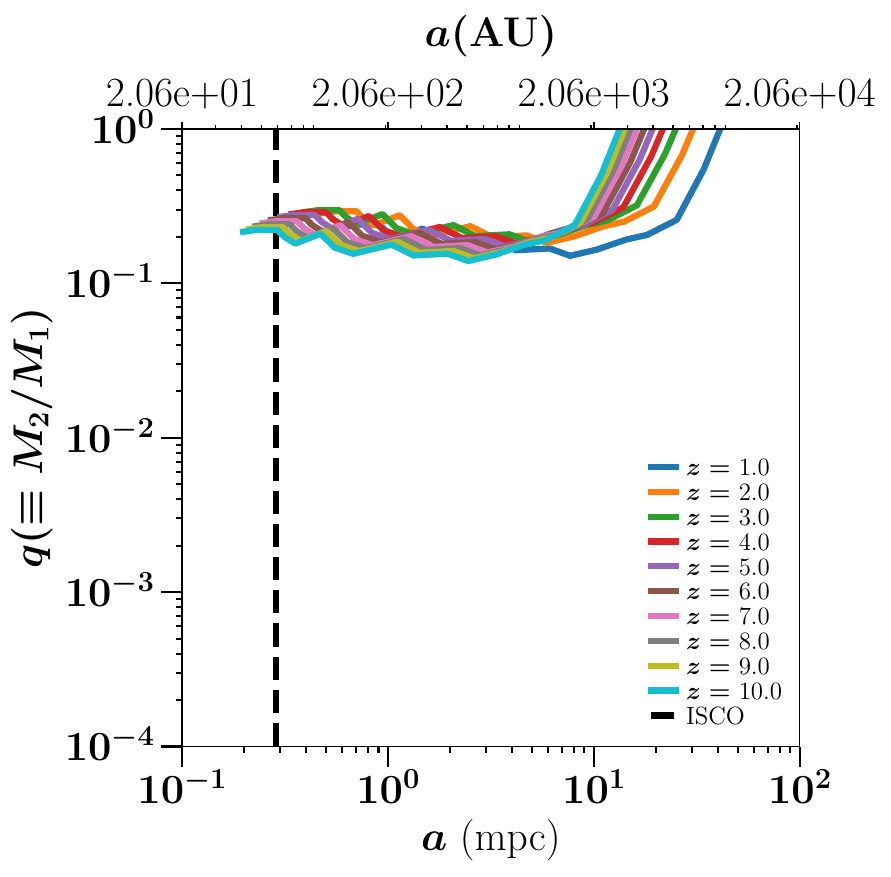}
\end{center}
\caption{Constraints on the $q-a$ plane from the SKA sensitivity curve and assuming a primary black hole mass $M_1 = 10^9 M_{\odot}$ at various redshifts. The dashed line shows the innermost stable circular orbit (ISCO).}
\label{fig:9msunskaz}
\end{figure}

\begin{figure}
\begin{center}
\includegraphics[width =\columnwidth]{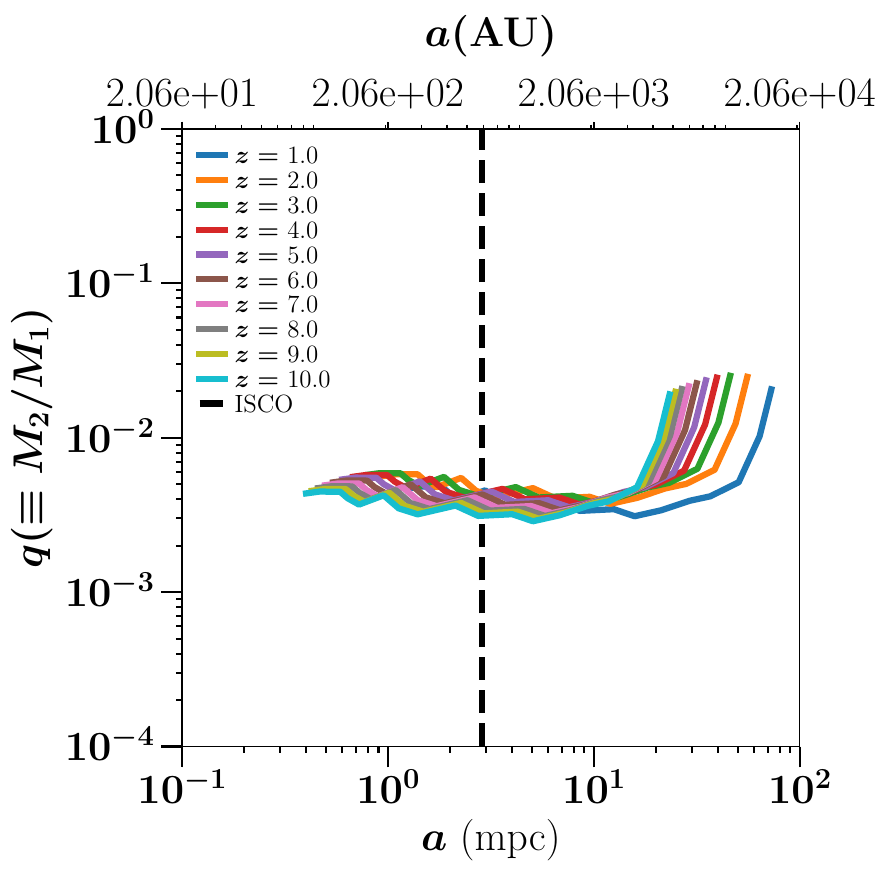}
\end{center}
\caption{Same as Fig. \ref{fig:9msunskaz}, but for the primary black hole mass $M_1 = 10^{10} M_{\odot}$.}
\label{fig:10msunskaz}
\end{figure}

The above equations can be used along with the SKA sensitivity curve   to forecast constraints on the separation and minimum mass ratio of the black holes. This is done by 
considering two fiducial  primary BH masses, $M_1 = 10^9$ and $10^{10} M_{\odot}$, assumed to be located at a range of redshifts between $z \sim 1$ and $z \sim 10$. We solve \eq{freqstrain} to derive the lower limits on the mass ratio ($q = M_2/M_1$) for both cases, as a function of separation $a$, allowing the the strain versus frequency curve to be translated into the $q-a$ plane as  
shown in Figs. \ref{fig:9msunskaz}, \ref{fig:10msunskaz}. 

{ A few comments on the qualitative behaviour of the curves in the $q-a$ plane are in order. \eq{freqstrain} implies the relation $\mathcal{M}_{\rm obs} \propto h d_L/f_{\rm obs}^{2/3}$, which can be approximated (in the regime $M_2 \ll M_1$) as  $M_2 \propto h d_L/(f_{\rm obs}^{2/3} (1 + z)^{5/3})$. Hence, for a given strain $h$ and separation $a$, we obtain $M_2 \propto d_L/(1+z)$. For a fixed $f_{\rm obs}$, $a$ is fixed up to a $(1+z)$ factor in the regime $M_2 \ll M_1$. The $(1+z)$ effect shows up as the linear shift in curves (at fixed $q$) as $z$ is varied (the shift is by a factor of $\sim 5$ from the highest to the lowest redshift). This leads to a very modest change (by a factor of order $\sim 3$) in $q$ as $z$ varies from 1 to 10, as can be seen from Figs. \ref{fig:9msunskaz},\ref{fig:10msunskaz}.
Thus, the range of secondary black hole masses constrainable by the SKA varies only mildly with $a$ at all redshifts, with the minimum mass ratio $q_{\rm min} \sim \{0.25, 0.005\}$ for $M_{\rm 1} = \{10^9, 10^{10}\} M_{\odot}$ respectively.} The other important finding is that primary black hole masses less than $M_1 \sim 10^9 M_{\odot}$ are undetectable by the SKA at all redshifts $z \gtrsim 1$.
Of the range in redshifts, the most relevant for unravelling the growth history of black holes is that close to the highest redshift quasars, whose electromagnetic counterparts may allow for robust determination of the source properties.

\begin{figure}
\begin{center}
\includegraphics[width =\columnwidth]{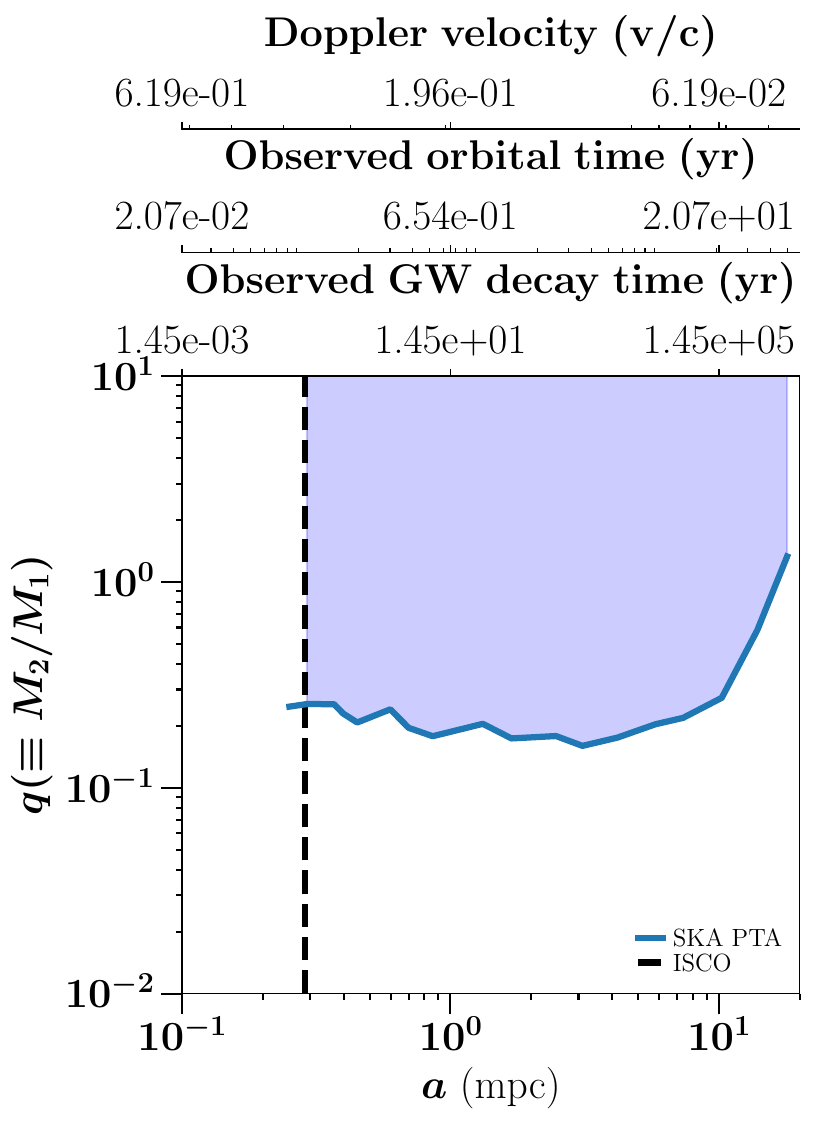}
\end{center}
\caption{Constraints on the $q-a$ plane from the SKA sensitivity curve and assuming a primary black hole mass $M_1 = 10^9 M_{\odot}$ at $z \sim 6.8$. The dashed line shows the innermost stable circular orbit (ISCO), while the shaded region indicates the regime of detectability of the binary SMBH with SKA PTA The top axes show the observed gravitational wave decay time and the orbital period  of the binary, and the velocity of the lower-mass black hole.  The range of velocities in the regime of detectability is of the order of $0.06c - 0.2c$, ensuring good prospects for prompt electromagnetic follow-up through Doppler boosting of the light curves.}
\label{fig:9msunska}
\end{figure}

\begin{figure}
\begin{center}
\includegraphics[width =\columnwidth]{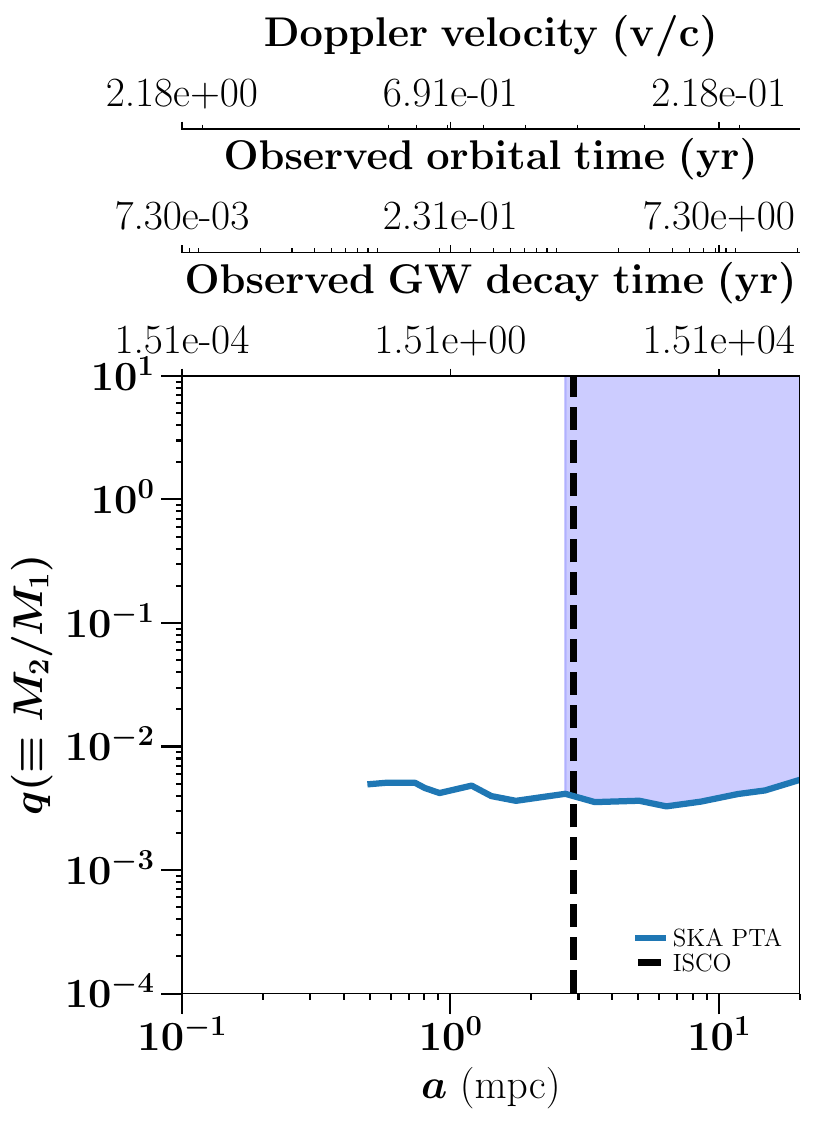}
\end{center}
\caption{Same as Fig. \ref{fig:9msunska}, for a primary black hole of mass $M_1 = 10^{10} M_{\odot}$ at $z \sim 6.8$.}
\label{fig:10msunska}
\end{figure}

We now replot the constraints on the $q-a$ plane obtained from the primary black hole assumed to be located at $z \sim 6.8$, the redshift of the most luminous quasar detected so far.  
This is shown in Figs. \ref{fig:9msunska} and \ref{fig:10msunska}.  In each case, the observed orbital time period of the binary for the preferred mass ratio ($q = 0.25$ and $q = 0.005$ respectively) is indicated as a function of separation:
\begin{equation}
t_{\rm orbit, obs} = \frac{2}{f_{\rm obs}}
\label{torbitobs}
\end{equation}
Also shown is the observed gravitational wave emission (or binary decay) time, 
\begin{equation}
t_{\rm gw, obs} = \frac{5}{256} \frac{c^5 a^4 (1+z)}{G^3 M_1 M_2 (M_1 + M_2)}
\label{tgwobs}
\end{equation}
and the velocity of the lower-mass black hole \citep[in units of $c$, assuming a circular orbit, e.g. ][]{dorazio2015a}:
\begin{equation}
v = \left(\frac{2 \pi}{1 + q}\right)\left(\frac{G M_{\rm BH}}{4 \pi^2 t_{\rm orbit, rest}}\right)^{1/3}
\end{equation}
and $t_{\rm orbit, rest}$ is derived from \eq{torbitobs} by dividing by $(1+z)$.

We can calculate the number of binary black holes expected at $z \sim 6$ as a function of mass ratio  by using the formalism for the merger rate of dark matter halos per unit 
redshift ($z$) and halo mass fraction ($\xi$), as formulated by \citet{fakhouri2010}:
\begin{eqnarray}
\frac{d n_{\rm halo}}{dz d\xi} &=& A \left(\frac{M_{\rm h}}{10^{12} 
M_{\odot}}\right)^{\alpha} \xi^{\beta} 
\exp\left[\left(\frac{\xi}{\bar{\xi}}\right)^{\gamma_1}\right] (1+ z)^{\eta}
\label{halomerger_rate}
\end{eqnarray}
where $M_{\rm h}$ is the halo mass.  In the above equation, { the best-fitting parameters constructed from the halo merger trees
have the values}  $\alpha = 0.133$, $\beta = -1.995$,
$\gamma_1 = 0.263$, $\eta = 0.0993$, $A = 0.0104$ and $\bar{\xi} = 9.72 \times 
10^{-3}$.
Substituting the empirically constrained black hole-halo mass relation \citep[e.g.,][]{wyithe2002}:
\begin{equation}
 M_{\rm BH} = M_{\rm h} \epsilon_0 \left(\frac{M}{10^{12} M_{\odot}}\right)^{\gamma/3 - 
1} \left(\frac{\Delta_v \Omega_m h^2}{18 \pi^2}\right)^{\gamma/6} 
(1+z)^{\gamma/2} \, ,
\label{mbhmhalo}
\end{equation} 
in which $\log_{10} \epsilon_0 = -5.02 $, $\gamma = 4.53 $ \citep[e.g.,][]{hploeblisa2020},
we derive:
\begin{eqnarray}
&& \frac{d N_{\rm BH,mrg}}{dz dq}  = A_1 f_{\rm bh}
\left(\frac{M_{\rm h}}{10^{12} M_{\odot}} \right)^{\alpha} \nonumber \\
&& \  \times q^{3/\gamma - 1 +3\beta/\gamma} 
(1+z)^{\eta} \exp\left[\left(\frac{q}{\bar{q}}\right)^{3\gamma_1/\gamma}\right] 
 \label{bhmergerrate}
\end{eqnarray}
In the above expression,  $f_{\rm bh}$ denotes the fraction of dark matter haloes occupied by black holes, and $q$ is the black hole mass ratio, and is related to $\xi$ as $q = \xi^{\gamma/3}$, $\bar{q}(\gamma) = 
\bar{\xi}^{\gamma/3}$ and $A_1 = (3/\gamma) A$. 
From this, we can define the number of binary black hole mergers expected per unit comoving volume as:
\begin{eqnarray}
&& \frac{d n_{\rm BHB}}{dz dq d \log_{10} M_{\rm BH}} =  A_1 f_{\rm bh} \frac{3}{\gamma} 
\left(\frac{M_{\rm h}(M_{\rm BH})}{10^{12} M_{\odot}} \right)^{\alpha} \nonumber \\
&& \  \times q^{3/\gamma - 1 +3\beta/\gamma} 
(1+z)^{\eta} \exp\left[\left(\frac{q}{\bar{q}}\right)^{3\gamma_1/\gamma}\right] \frac{dn_{\rm h}}{d \log_{10} M_{\rm h}}
 \label{bhnumberdensity}
\end{eqnarray}
in which the $dn_{\rm h}/d \log_{10} M_h$ term represents the dark matter halo mass function (taken to follow  \citet{sheth2002}) and $M_{\rm h}(M_{\rm BH})$ denotes the dark matter halo mass expressed as a function of black hole mass, obtained by inverting \eq{mbhmhalo}.
The number density of gravitational wave events can now be found by multiplying the above expression by $t_{\rm gw} (a_{\rm gr})(dz/dt)$, where $t_{\rm gw}(a_{\rm gr})$ is the gravitational wave decay timescale in the rest frame of the binary, which for masses $M_1$ and $M_2$ is given by \eq{tgwobs} divided by $(1+z)$.
It is evaluated at the scale $a_{\rm gr}$ --  the radius at which emission of gravitational waves takes over as the dominant channel  of coalescence -- which is defined by:
\begin{equation}
a_{\rm gr} = A | {\rm ln} A|^{0.4} a_{\rm h}
\end{equation}
with
\begin{equation}
A = 9.85 \left(\frac{M_1}{M_2}\right)^{0.2} \left(\frac{M_1 + M_2}{2 M_2}\right)^{0.4} \left(\frac{\sigma}{c}\right)
\end{equation}
and where
\begin{equation}
a_{\rm h} = 2.8 \left(\frac{M_2}{10^8 M_{\odot}}\right)  \left(\frac{\sigma}{200 {\rm km/s}}\right)^{-2} {\rm pc}
\end{equation}
is the hardening scale of the binary \citep[e.g.,][]{merritt2000,kulkarni2012}.  In both of the above expressions, $\sigma$ represents the one-dimensional velocity dispersion of the primary host galaxy, which is related to its black hole mass by \citep{tremaine2002}:
\begin{equation}
\sigma = 208 \ {\rm km/s} \ \left(\frac{M_1}{1.56 \times 10^8 M_{\odot}}\right)^{1/\gamma}
\end{equation}
with $\gamma$ as  defined in \eq{mbhmhalo}.
For each of the candidate primary black hole masses above, we can integrate the expression over the requisite $q$-range to find the total number of binary SMBHs $\phi_{\rm BHB, gw}   (M_{\rm BH}, q_{\rm min})$ expected per unit comoving volume.  Assuming the binary merger proceeds to coalescence without delay \citep{fang2022}, we find
\begin{eqnarray}
&& \phi_{\rm BHB, gw}   (M_{\rm BH}, q_{\rm min}) \equiv \frac{dn_{\rm BHB}}{d \log_{10} M_{\rm BH}} |_{q_{\rm min}}  \nonumber \\
&=& \int_{q_{\rm min}}^{1} dq \frac{d n_{\rm BHB}}{dt \ dq \ d \log_{10} M_{\rm BH}} t_{\rm gw} (a_{\rm gr})
\label{phibhb}
\end{eqnarray}
over the $q$-range starting from the minimum value of $q$ under consideration. { In the case of the SKA, for example, $q_{\rm min} \sim 0.25$ when $M_{\rm BH} = 10^9 M_{\odot}$ and $q_{\rm min} \sim 0.005$ when $M_{\rm BH} = 10^{10} M_{\odot}$, fairly independently of separation as shown in Fig. \ref{fig:9msunskaz}.}

From \eq{phibhb}, one obtains the total number of black hole binaries in a given redshift interval:
\begin{eqnarray}
\frac{d N_{\rm BHB, gw}(M_{\rm BH}, q_{\rm min})}{d \log_{10} M_{\rm BH}} = dV(z_1, z_2) 
  \  \phi_{\rm BHB, gw} (M_{\rm BH}, q_{\rm min})
\label{nbhbgw}
\end{eqnarray}
where $d V$ represents the comoving volume under consideration. All of the above expressions are generic and valid for the whole range of redshift and primary black hole mass values.
In particular, they can be used to compute the number of the number of binaries detectable by the SKA PTA around $z \sim 6.8$ and over the interval $6 < z < 7.1$, for both the primary black hole masses considered earlier {along with their corresponding $q_{\rm min}$ values}.  This is indicated on the third column of  Table \ref{table:binarybhnumbers}. The quantities are also a function of the solid angle on the sky  area probed by the survey (we quote results for the full sky in Table \ref{table:binarybhnumbers}; results for a partial sky coverage with solid angle $\Omega$ are obtained by multiplying the values by $\Omega/4 \pi$).

\begin{table*}
 \begin{center}
 \begin{tabular}{cccc}
 \hline
  $M_{\rm BH}$ & $\phi_{\rm BHB, gw}$ (cMpc$^{-3}$) & $dN_{\rm BHB, gw}/d \log_{10} M_{\rm BH}$ & $n_{\rm gal, LSST}$  (deg$^{-2}$) \\
  \hline \\
  $10^9 M_{\odot}$ & $4.2 \times 10^{-4}$ & $1.0 \times 10^8$ & $\sim 3600$  \\
  \\
  $10^{10} M_{\odot}$ & $1.54 \times 10^{-5}$  & $2.4 \times 10^6 $ & $\sim 100$  \\
  \hline
 \end{tabular}
 \end{center}
 \caption{ The table shows the expected numbers of black hole binaries detectable by the SKA PTA  at the median redshift of $z \sim 6.8$, and their galaxy counterparts detectable by LSST. The first column shows the mass of the primary black hole. The second column indicates the number of binaries expected per unit comoving volume, which is obtained from \eq{phibhb} for each of the primary black hole masses in column 1. The third column shows the total number of detectable binaries in the full sky over a redshift interval $6 < z < 7.1$ per unit logarithmic range of black hole mass. The last column indicates the number of electromagnetic counterpart galaxies expected by LSST per square degree of sky area \citep{lsst2009}.}
 \label{table:binarybhnumbers}
\end{table*}

\section{Electromagnetic counterparts}
\label{sec:counterparts}

Once a gravitational wave event has been detected,  a number of methods may be used to locate its electromagnetic counterpart,  such as  frequency matching \citep{xin2021,  schneider2021} of the periodicity of the quasar optical lightcurves \citep{graham2015, millon2022, hayasaki2016, charisi2016, haiman2017} or spectroscopic detection of Doppler shifts that modulate the observed flux \citep{Bogdanovic2011, dorazio2018, dorazio2013, dorazio2015} which have already been used to identify candidate supermassive binaries such as OJ 287 \citep{Valtonen_2016}.  As we have seen above,  this is especially convenient for the binaries within the SKA PTA regime,  since their observed orbital timescales are of the order of a few weeks to years, and Doppler velocities a few hundredths to tenths of $c$ (indicated on Figs. \ref{fig:9msunska}, \ref{fig:10msunska}).  For the remainder of the analysis,  therefore, we assume that the counterpart identification proceeds via one of the methods listed above.

It is forecasted that SKA PTA may be able to localize a region of area  of about $70$ deg$^2$ in an optimistic scenario \citep{wang2017}. We use this along with the previous findings to derive the expected number of electromagnetic counterparts (galaxies and quasars) as a  function of the primary black hole mass and other parameters.

\subsection{Galaxy counterparts with LSST}
\label{sec:galaxies}

The LSST on the Rubin Observatory can be used to identify the  galaxies hosting high-redshift black holes. To estimate the number of such counterparts, we  use a procedure similar to that followed in \citet{hploeblisa2020} for analogous counterparts of LISA gravitational wave detections. Given the large uncertainties in galaxy number counts at $z \sim 6$ and beyond, we only quote approximate values here. 
We start with the connection between the $g-r$ color and the galaxy stellar mass-to-light ratio, given  by \citep{wei2020}:
\begin{equation}
\log_{10} (M_*/L_z) = a_z + b_z (g-r)
\end{equation}
with $a_z = -0.731, b_z = 1.128$ for the $z$-band, which is assumed to hold out to the redshifts under consideration.
We adopt a typical value of $g-r = 1$ for LSST galaxies \citep{lsst2009}. Using the above relation along with the 
 black hole - galaxy bulge mass relation to determine the stellar mass, we connect the black hole masses to the observed $L_z$ luminosities and absolute magnitudes to calculate the resultant $z$-band luminosity of the merged galaxy. We find that for a primary black hole mass of $M_{\rm BH} = 10^{10} M_{\odot}$ at $z \sim 6.8$, the apparent magnitude of the galaxy in the $z$-band is 25.6, which leads to a number density of $\sim 10^{-1.5}$ galaxies per square arcmin \citep{bouwens2006, lsst2009}. This implies that in the SKA localization ellipse of 70 deg$^2$,  we expect about 8000  electromagnetic counterparts of such a system with LSST.  A similar analysis for a $10^9 M_{\odot}$ black hole leads to  a $z$-band magnitude of 27.8, which translates into 1 galaxy arcmin$^{-2}$, or $\sim 2.5 \times 10^5$ galaxies within the SKA localization ellipse.  In both cases, if the candidate host galaxy from LSST is not uniquely identifiable and the quasar is not dormant, we would need to turn to possible quasar counterparts.

\begin{figure}
\begin{center}
\includegraphics[width =\columnwidth]{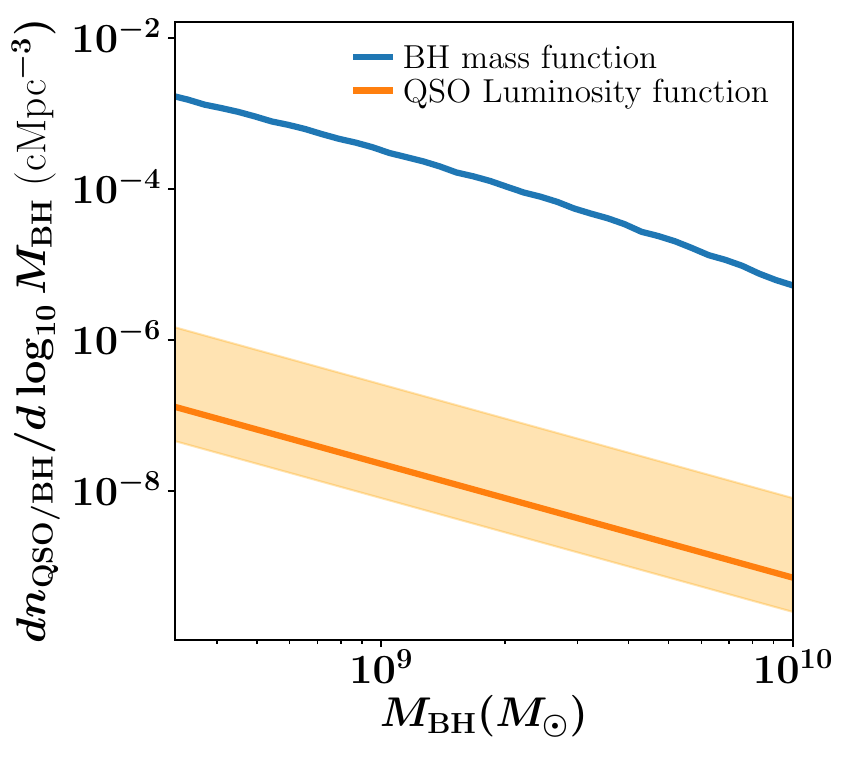}
\end{center}
\caption{The mass function of active black holes at  $z \sim 6.8$ computed from the quasar luminosity function (orange), compared to the empirically derived mass function of all black holes at this redshift (blue). The solid orange line represents an Eddington ratio of $ \eta_{\rm Edd} = 0.5$ and the shaded region covers the range $0.1 \leq \eta_{\rm Edd} \leq 1$, which is motivated by observations at $z \simeq 6$.}
\label{fig:qsolumfunc}
\end{figure}

\begin{figure}
\begin{center}
\includegraphics[width =\columnwidth]{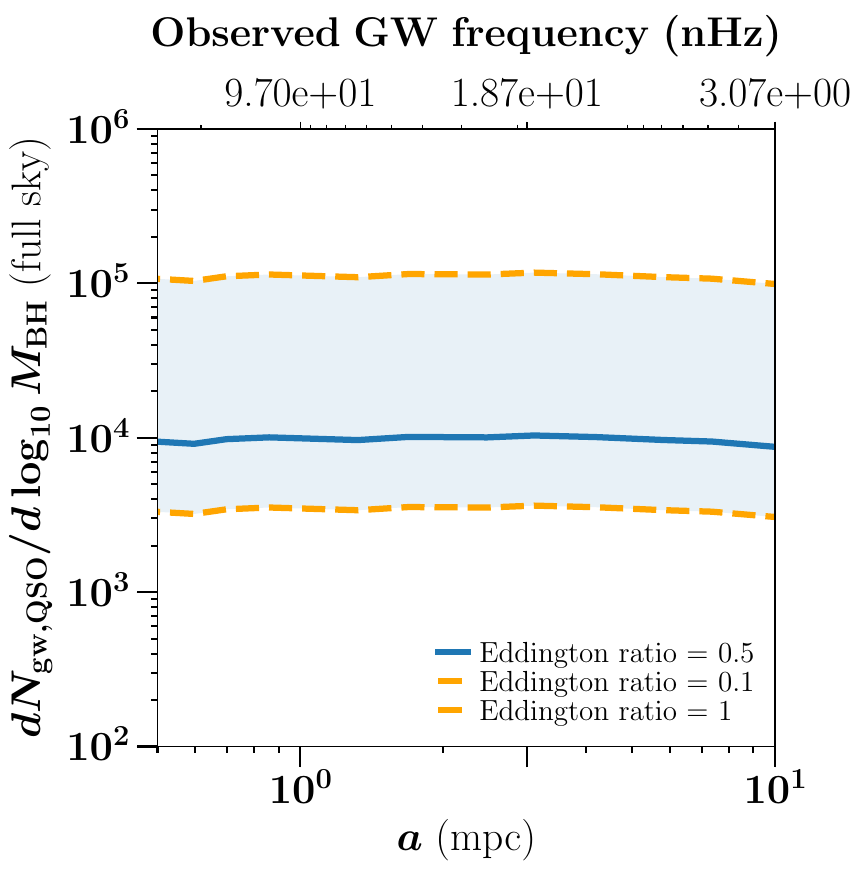}
\end{center}
\caption{Expected number of active quasars at $z \sim 6.8$ over the whole sky as a function of separation, with the primary black hole mass $M_{\rm BH} = 10^{9} M_{\odot}$. The observed frequency corresponding to the separation is shown on the top axis. }
\label{fig:avsnqso9}
\end{figure}

\subsection{Quasar counterparts}
\label{sec:qsos}

We now summarize prospects for uniquely identifying the quasar counterpart (even though these are much less numerous) to the binary merger. To do this, we estimate the number of active quasars at this redshift based on existing survey constraints. We use the latest observational compilations of the quasar luminosity function at $z \sim 0-7$ \citep{shen2020} that advocates modelling the luminosity function as a single power law at the highest redshifts:

\begin{equation}
\phi(L) \equiv \frac{dn_{\rm QSO}}{d \log_{10} L} = \frac{\phi_*}{2(L/L_*)^{\gamma_1}}
\end{equation}
This measures the number density of quasars, $n_{\rm QSO}$ per unit logarithmic luminosity interval, with the best-fitting parameters $\log \phi_*  = -5.452, \ \log \ (L_*/L_{\odot}) = 11.978, \ \gamma_1 = 1.509$.

Treated as a function of the host black hole mass, the above function can be re-expressed as the mass function of the active black holes  hosting the quasars by introducing the Eddington ratio, $\eta_{\rm Edd} \equiv L/L_{\rm Edd}$, as:
\begin{eqnarray}
\phi(M_{\rm BH}| {\rm QSO}, \eta_{\rm Edd}) &=& \frac{dn_{\rm QSO}}{d \log_{10} M_{\rm BH|QSO}} \nonumber \\
 &=&  \phi(\eta_{\rm Edd} L_{\rm Edd}(M_{\rm BH})) \left|\frac{d \log_{10} L}{d \log_{10}  M_{\rm BH}}\right| \nonumber \\
 \label{blackholemfqso}
\end{eqnarray}
in which $L_{\rm Edd}(M_{\rm BH}) = 1.3 \times 10^{38} (M_{\rm BH}/M_{\odot}) L_{\odot}$ is the Eddington luminosity.
If $\eta_{\rm Edd}$ is treated as a constant independent of $M_{\rm BH}$ -- as assumption we make for the remainder of the analysis -- the second term goes to unity and the above relation reduces simply to $\phi(\eta_{\rm Edd} L_{\rm Edd}(M_{\rm BH}))$.

We now compare this to the black hole mass function of all galaxies (i.e. irrespective of the activity of the quasar). This is  computed from the best-fitting black hole mass - halo mass relation in \eq{mbhmhalo}:
\begin{eqnarray}
\phi_{\rm BH} (M_{\rm BH}) &\equiv& \frac{d n_{\rm BH}}{d \log_{10} M_{\rm BH}} \nonumber \\
&=& f_{\rm BH} \frac{d n_{\rm h}}{d \log_{10} M_{\rm h}} \left|\frac{d \log_{10} M_{\rm h}}{d \log_{10}  M_{\rm BH}}\right|
\label{blackholemfall}
\end{eqnarray} 
noting that $\left|d \log_{10} M_{\rm h}/d \log_{10}  M_{\rm BH}\right| =  3/\gamma$, with $\gamma$ describing the slope of $M_{\rm BH} - M$ relation developed in \eq{mbhmhalo}.  For the redshifts under consideration, both  observations and theoretical arguments  \citep{ricarte2018} suggest that the occupation fraction $f_{\rm BH} \approx 1$, i.e. each halo is expected to host a black hole. This leads to the blue curve shown in Fig. \ref{fig:qsolumfunc}, which is also consistent with observational estimates from galaxy luminosity functions \citep{Gallo_2019}. The quasar luminosity function is shown by the orange line in Fig. \ref{fig:qsolumfunc}, with the shaded area enclosing the allowed range of Eddington ratios $0.1 \leq \eta \leq 1$ as suggested by  observations of quasars above $z \sim 5.8$ \citep{onoue2019,  lusso2022}.

The ratio of \eq{blackholemfqso} to \eq{blackholemfall} defines the \textit{active} fraction of black holes \citep[e.g.,][]{shankar2013} as a function of their mass: 
\begin{equation}
f_{\rm active} (M_{\rm BH}|\eta_{\rm Edd}) = \phi(M_{\rm BH}|{\rm QSO}, \eta_{\rm Edd})/\phi_{\rm BH} (M_{\rm BH})
\label{factive}
\end{equation}
This number is of the order of  
$10^{-4} - 10^{-5}$ depending on the assumed mass and Eddington ratio of the black holes.
It is provided in the second column of Table \ref{table:qsobinarybhnumbers} for the two masses of the black holes under present consideration and at $\eta = 0.5$.

\begin{table}
 \begin{center}
 \begin{tabular}{ccc}
 \hline
  $M_{\rm BH}$ & $f_{\rm active}$ ($\eta = 0.5$) & $N_{\rm gw, QSO}$  (in 70 deg$^{-2}$)  \\
  \hline \\
  $10^9 M_{\odot}$ & $6.8 \times 10^{-5}$ &  $\sim 7-45$ \\
  \\
  $10^{10} M_{\odot}$ & $1.3 \times 10^{-4}$  & $\lesssim 1$ \\
  \hline
 \end{tabular}
 \end{center}
 \caption{The active quasar fraction as a function of primary black hole mass at a median redshift of $z \sim 6.8$ (second column), and the total number of quasar counterparts  over a redshift interval $6 < z < 7.1$ (third column) expected inside the SKA localization ellipse, assuming a logarithmic mass interval of $d \log_{10} M_{\rm BH} = 0.1$ around its central value.}
 \label{table:qsobinarybhnumbers}
\end{table}

\begin{figure}
\begin{center}
\includegraphics[width =\columnwidth]{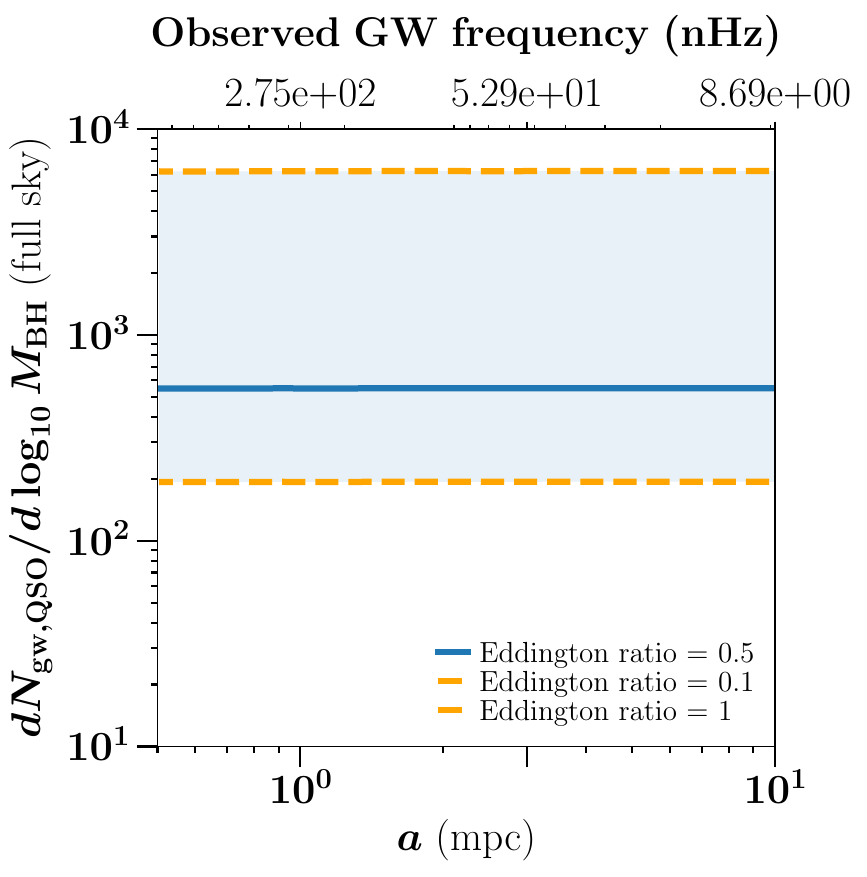}
\end{center}
\caption{Same as Fig. \ref{fig:avsnqso9}, for a primary black hole of mass $M_{\rm BH} = 10^{10} M_{\odot}$.}
\label{fig:avsnqso10}
\end{figure}

\begin{figure}
\begin{center}
\includegraphics[width =\columnwidth]{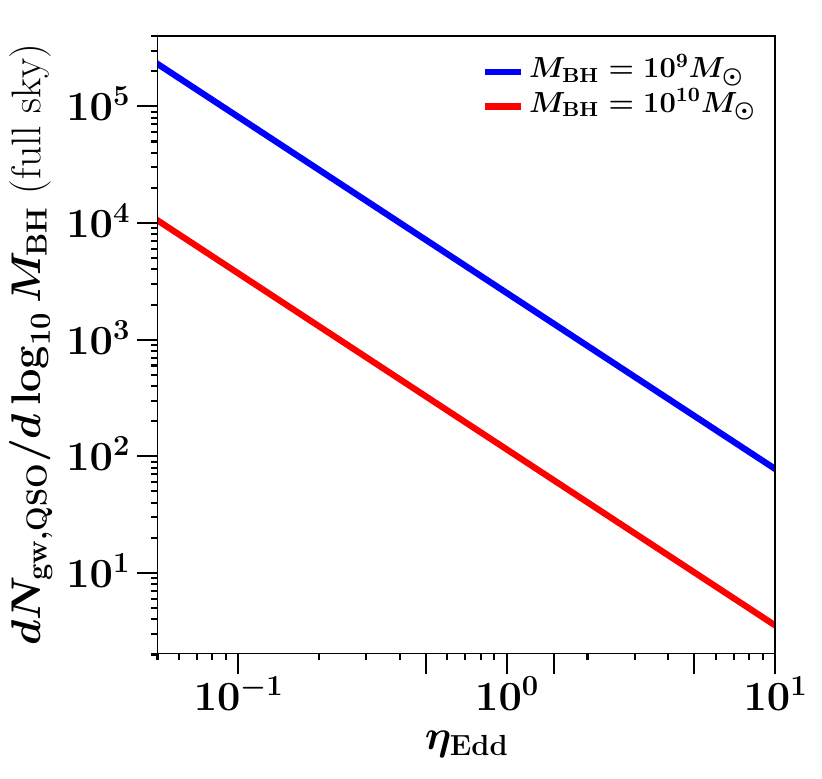}
\end{center}
\caption{Expected number of active quasar counterparts of SMBH binary mergers with $M_{\rm BH} = 10^9, 10^{10} M_{\odot}$ at $z \sim 6.8$ over the whole sky, as a function of the Eddington ratio. The separation in both cases is fixed to $a = 10$ mpc.}
\label{fig:etanqso}
\end{figure}

We can now calculate the  number of active quasar counterparts
 as a function of the black hole separation (or equivalently, the observed frequency of the gravitational wave event) by combining the quasar luminosity function with the detectable region in Figs. \ref{fig:9msunska}, \ref{fig:10msunska}. This is done by computing the quantity
\begin{eqnarray}
& & dN_{\rm gw,QSO} (M_{\rm BH}, q_{\rm min}|\eta_{\rm Edd}) = f_{\rm active}(M_{\rm BH}|\eta_{\rm Edd}) \nonumber \\
 &\times & dN_{\rm BHB, gw} (M_{\rm BH}, q_{\rm min})
 \label{nqsotonbhb}
\end{eqnarray}
where $f_{\rm active}$  and $dN_{\rm BHB, gw}$ are defined in \eq{factive} and \eq{nbhbgw} respectively. Figs. \ref{fig:avsnqso9} and \ref{fig:avsnqso10} show this quantity per unit logarithmic black hole mass interval at $z \sim 6.8$, for  both the masses of the primary black hole under consideration, as a function of the black hole separation $a$ (which determines $q_{\rm min}$ according to Figs. \ref{fig:9msunska}, \ref{fig:10msunska}). In  both plots, the comoving volume is taken to be that enclosed between $z \sim 6$ and $z \sim 7.1$ with a full-sky areal coverage. The near-constancy of $dN_{\rm QSO}/d \log_{10} M_{\rm BH}$ with separation is a consequence of $q_{\rm min}$ being nearly independent of $a$, as seen in previous figures. Note, however, that the number of active quasars expected is strongly dependent on their mean Eddington ratio -- this is  shown by plotting $dN_{\rm QSO}/d \log_{10} M_{\rm BH}$  as a function of $\eta_{\rm Edd}$ at  a fiducial separation of  $a = 10$ mpc in Fig. \ref{fig:etanqso}, for both values of $M_{\rm BH}$ under consideration.

\begin{figure}
\begin{center}
\includegraphics[width =\columnwidth]{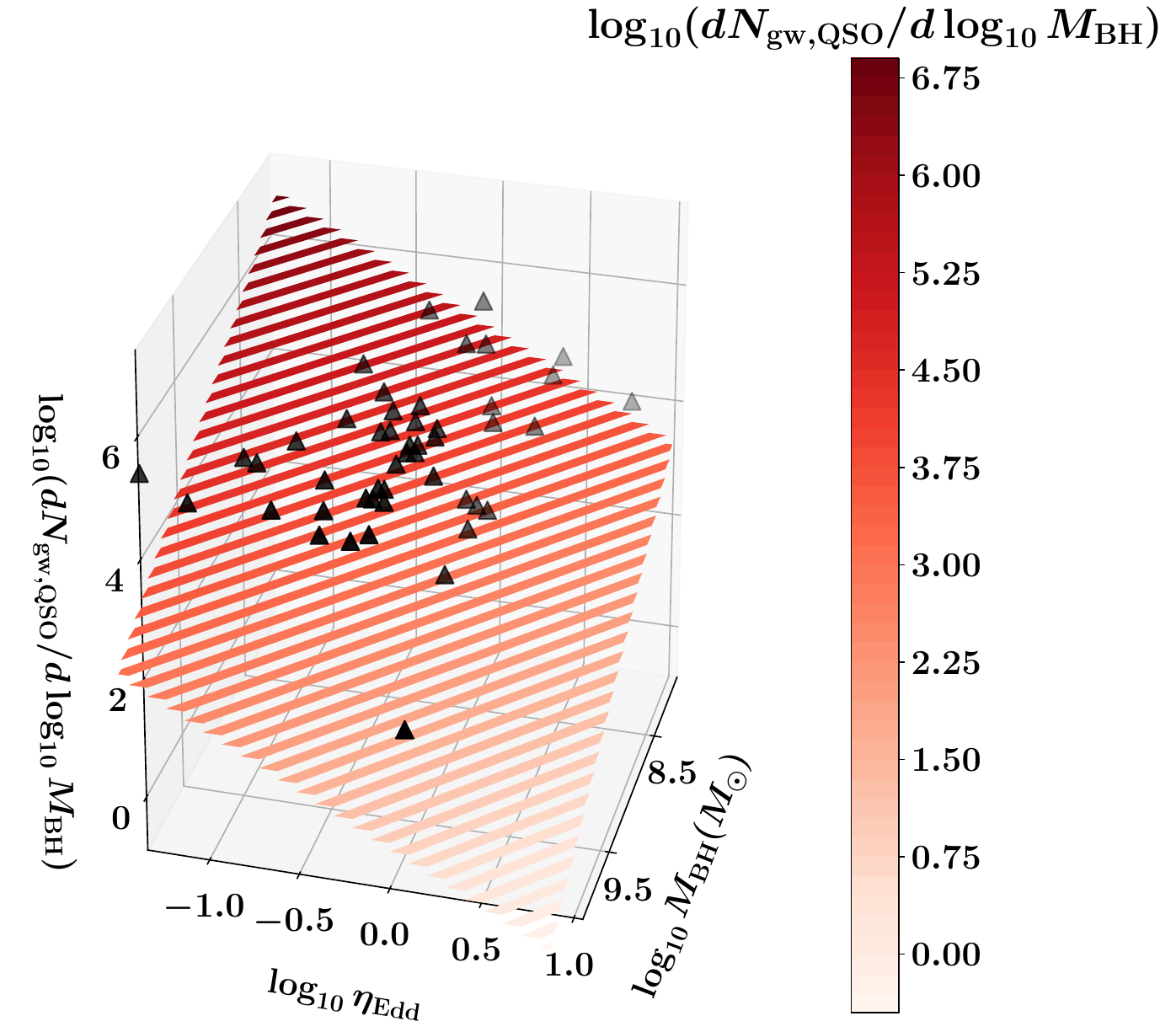}
\end{center}
\caption{Predicted number of quasar counterparts (in the full sky) for black holes at $z \gtrsim 6$, as a joint function of their masses and Eddington ratios. Black holes with known masses and Eddington ratios from the literature \citep{kim2019} are indicated by the black triangles superimposed on the figure.}
\label{fig:3dqsos}
\end{figure}

The preceding analysis thus enables us to connect the number of high-redshift gravitational wave counterparts detected to an important property of the quasar, namely its accretion efficiency. We can now combine these findings with observations of $z > 6$ quasars in the literature for which mass and  Eddington ratio estimates already exist. We use the compilation in \citet{kim2019} which lists the derived rest-UV and FIR properties of redshift $z \gtrsim 6$ quasars.  In each case, we calculate the number of expected counterparts within the whole sky area if the system were in a binary. This is tabulated in Table \ref{table:qsoobs} and  shown in Fig. \ref{fig:3dqsos}, which also provides a three-dimensional representation of the results in the previous Fig. \ref{fig:etanqso}, and is obtained by gridded interpolation over both the mass and Eddington ratio ranges. Also shown in Table \ref{table:qsoobs} are the derived $i$-band magnitudes of the quasars, which can be estimated from the black hole masses and Eddington ratios \citep{haiman2009} by adopting a typical quasar spectral energy distribution:
\begin{equation}
m_i = 24 \ + \ 2.5 \log \left[\left(\frac{\eta_{\rm Edd}}{0.3}\right) \left(\frac{M_{\rm BH}}{3 \times 10^6 M_{\odot}}\right)^{-1} \left(\frac{d_{L}(z)}{d_L(z = 2)}\right)^2\right]
\end{equation}
{ A possible source of error in the magnitudes above may arise from the correction to $d_L$ induced by weak lensing of the GWs through
the large-scale inhomogeneities  along their path \citep{hirata2010}. Assuming this correction  to be given  by the fitting form introduced in that work, it is expected to lead to about a 10-20\% error in the inferred magnitude.\footnote{ While the said fitting form is calibrated until $z \sim 3$, for the purposes of this discussion, we assume it to be valid out to $z \sim 6$.} Another contribution to the uncertainty stems from the redshift determination of the source, which we assume to be negligible for the present purposes, being accomplished by an independent high-resolution measurement (such as the UV/FIR surveys used for the results in Table \ref{table:qsoobs}).}

\begin{table*}
\begin{tabular}{|c|c|c|c|c|c|}
\hline 
& & &  & & \\
Name & $M_{\rm BH}$ ($10^8 M_{\odot})$ & $z$ & $\eta_{\mathrm{Edd}}$ & $dN_{\mathrm{gw, QSO}}/d \log_{10} M_{\rm BH}$ & $m_i$ \\
& & & & &  \\
\hline 
 & & & & & \\
J0005-0006 & 0.8 & 5.844 & 5.5 & 6160.0 & 26.0 \\
J0028+0457 & 28.8 & 5.99 & 0.29 & 41140.0 & 19.0 \\
J0033-0125 & 26.3 & 6.02 & 0.1 & 54110.0 & 18.0 \\
J0050+3445 & 25.7 & 6.253 & 0.55 & 31810.0 & 20.0 \\
J0055+0146 & 2.4 & 5.983 & 1.07 & 197070.0 & 23.0 \\
J0100+2802 & 107.2 & 6.3 & 1.2 & 60.0 & 20.0 \\
J0109-3047 & 13.5 & 6.763 & 0.3 & 118070.0 & 21.0 \\
J0136+0226 & 3.1 & 6.21 & 0.55 & 875690.0 & 23.0 \\
J0210-0456 & 0.8 & 6.438 & 2.19 & 26050.0 & 26.0 \\
J0221-0802 & 6.9 & 6.161 & 0.3 & 317130.0 & 21.0 \\
J036.5078+03.0498 & 30.2 & 6.533 & 0.54 & 28060.0 & 20.0 \\
J0227-0605 & 1.8 & 6.21 & 0.95 & 638920.0 & 24.0 \\
J0303-0019 & 3.3 & 6.079 & 0.91 & 316340.0 & 23.0 \\
J0305-3150 & 9.1 & 6.61 & 0.68 & 137820.0 & 22.0 \\
J0353+0104 & 15.8 & 6.072 & 0.68 & 57800.0 & 21.0 \\
J0836+0054 & 11.0 & 5.81 & 0.51 & 139390.0 & 21.0 \\
J0841+2905 & 10.0 & 5.95 & 0.87 & 73890.0 & 22.0 \\
J0842+1218 & 19.1 & 6.069 & 0.63 & 50420.0 & 21.0 \\
J1030+0524 & 13.2 & 6.302 & 0.74 & 59000.0 & 21.0 \\
J1048+4637 & 19.1 & 6.198 & 1.17 & 910.0 & 21.0 \\
J167.6415-13.4960 & 3.0 & 6.505 & 1.26 & 13900.0 & 24.0 \\
J1120+0641 & 25.1 & 7.087 & 0.59 & 30270.0 & 21.0 \\
J1137+3549 & 52.5 & 6.01 & 0.33 & 15410.0 & 19.0 \\
J1148+5251 & 50.1 & 6.407 & 0.48 & 13150.0 & 19.0 \\
J1148+0702 & 14.8 & 6.34 & 0.74 & 53860.0 & 21.0 \\
J1205-0000 & 47.9 & 6.73 & 0.06 & 21940.0 & 17.0 \\
J1207+0630 & 44.7 & 6.03 & 0.17 & 22790.0 & 18.0 \\
J1250+3130 & 7.8 & 6.14 & 1.35 & 2880.0 & 23.0 \\
J1306+0356 & 11.0 & 6.017 & 0.87 & 62820.0 & 22.0 \\
J1335+3533 & 40.7 & 5.9 & 0.32 & 24480.0 & 19.0 \\
J1342+0928 & 7.8 & 7.527 & 1.55 & 2510.0 & 23.0 \\
J1411+1217 & 10.7 & 5.903 & 1.32 & 1980.0 & 22.0 \\
J1427+3312 & 7.6 & 6.12 & 1.17 & 3290.0 & 22.0 \\
J1429+5447 & 13.2 & 6.12 & 0.52 & 89510.0 & 21.0 \\
J1509-1749 & 29.5 & 6.121 & 0.62 & 25050.0 & 20.0 \\
J231.6576-20.8335 & 30.9 & 6.587 & 0.5 & 29150.0 & 20.0 \\
J1602+4228 & 15.5 & 6.08 & 0.95 & 26130.0 & 21.0 \\
J1623+3112 & 14.1 & 6.211 & 0.68 & 64020.0 & 21.0 \\
J1630+4012 & 11.0 & 6.058 & 0.68 & 103230.0 & 21.0 \\
J1641+3755 & 2.4 & 6.047 & 2.04 & 9900.0 & 24.0 \\
J2100-1715 & 9.3 & 6.087 & 0.43 & 203290.0 & 21.0 \\
J323.1382+12.2986 & 14.1 & 6.592 & 0.47 & 91730.0 & 21.0 \\
J2229+1457 & 1.2 & 6.152 & 2.14 & 24070.0 & 25.0 \\
J338.2298+29.5089 & 27.5 & 6.66 & 0.12 & 51190.0 & 19.0 \\
J2310+1855 & 43.7 & 5.96 & 0.58 & 14290.0 & 20.0 \\
J2329-0301 & 2.5 & 6.417 & 1.17 & 19120.0 & 24.0 \\
J2348-3054 & 20.4 & 6.902 & 0.18 & 83130.0 & 20.0 \\
J2356+0023 & 38.9 & 6.05 & 0.03 & 35350.0 & 17.0 \\
\hline
\end{tabular}
\caption{Predicted number of quasar counterparts and their $i$-band luminosities for black holes in the literature  for which masses and Eddington ratio estimates  exist, assuming the system to be in a binary. We use the compilation in \citet{kim2019} which includes the latest results from UV and FIR surveys of high-redshift quasars.}
\label{table:qsoobs}
\end{table*}

\begin{figure}
\begin{center}
\includegraphics[width =\columnwidth]{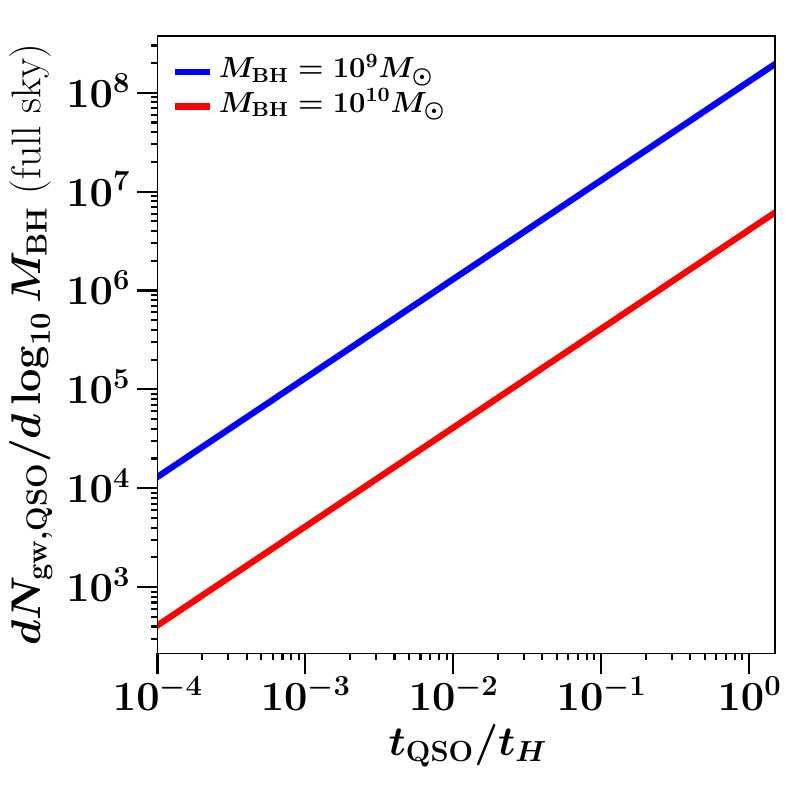}
\end{center}
\caption{Expected number of active quasar counterparts with $M_{\rm BH} = 10^9, 10^{10} M_{\odot}$ at $z \sim 6.8$ over the whole sky, as a function of the mean quasar age (expressed in terms of the Hubble time $t_H$).}
\label{fig:agenqso}
\end{figure}

The above arguments can also be reversed to place constraints on the active lifetime of quasars at these redshifts. This is done by not assuming an observationally determined quasar luminosity function as above -- but instead, \textit{defining} the expected active fraction in terms of the mean age of the quasars as:
\begin{equation}
f_{\rm active} = t_{\rm QSO}/t_H
\end{equation}
where $t_H$ is the Hubble time at the redshift under consideration. Using this for $f_{\rm active}$ in \eq{nqsotonbhb}, 
we plot the number of expected quasar counterparts as a function of the quasar age (in units of the Hubble time) in Fig. \ref{fig:agenqso} (with $q_{\rm min}$ assumed to be \{0.25, 0.005\} for the primary black hole mass $M_{\rm BH} = \{10^9, 10^{10}\} M_{\odot}$ respectively).

The above results can be converted into the number of active quasar counterparts within a sky area, as a function of the active lifetime of the quasar. This is shown for an area within the SKA localization ellipse of 70 deg$^2$ in the third column of Table \ref{table:qsobinarybhnumbers}.  It is found that the number of active quasar counterparts of $10^9 M_{\odot}$ black holes (in a logarithmic mass interval of $d \log_{10} M_{\rm BH} = 0.1$) within the SKA localization ellipse  is 
$N_{\rm gw, QSO} = 7-45$, and that for $10^{10} M_{\odot}$ black holes is $\lesssim 1$ ($N_{\rm gw, QSO} = 0.1-1.2$) for Eddington ratios in the range $\eta_{\rm Edd} = \{0.1, 1\}$. This -- as well as the findings of Table \ref{table:qsoobs} --  indicate that the quasar counterparts of the most massive black holes at $z \gtrsim 6$ can be uniquely identified within the SKA localization ellipse.

\section{Discussion}
\label{sec:discussion}

We have investigated  the ability of  SKA PTA to locate and identify SMBH binaries at $z \gtrsim 6$. We find that SKA PTA should be sensitive to black hole binaries with primary masses $M_{\rm BH} \gtrsim 10^9 M_{\odot}$ at all redshifts. The minimum mass ratio constrainable by the SKA varies from $\sim 0.005$ to $0.25$ as  $M_{\rm BH}$ goes from $10^9 M_{\odot}$ to $10^{10} M_{\odot}$, fairly independently of separation.    At $z > 6$, the regime of SKA PTA  detectability covers binary separations of about $\sim 1 - 50$ mpc, depending on the mass of the primary black hole.  Binaries with such  separations are expected to be close to final coalescence \citep{begelman1980, graham2015} with their orbital timescales of the order of a few weeks to years.

{ For the values of the black hole masses and separations considered here,  the gravitational wave decay timescale at the radius $a_{\rm gr}$ is of the order of 100 Myr, consistently with the results of simulations and analytical models \citep{tiede2020, loeb2007, nasim2020, bogdanovic2022}. We have assumed, following previous work \citep{hploeblisa2020} that the mergers proceed rapidly to coalescence. For small mass ratio mergers, it has been postulated that the dynamical friction timescale could be larger than the Hubble time \citep{callegari2011}, leading to the so-called ``final parsec problem'' \citep{yu2002, Milosavljevic2001}.  From observations of smaller-mass black hole binaries \citep{fishbach2021} at lower redshifts, this delay has been estimated as  a function of parameters such as the progenitor mass and metallicity, finding most binaries to have low delay times $< 500$ Myr. This number is likely to reduce further as the masses of the black holes become higher. There is also observational evidence from X-shaped radio sources \citep{dennett2002} favouring efficient binary coalescence \citep{merritt2002} which can, in turn, result from a number of processes, such as non-spherical galaxy shapes leading to filled loss cones, or gas-rich environments \citep{haiman2009, armitage2002, colpi2014}.  We leave a more detailed consideration of these effects to future work.}

Using empirically motivated constraints on the occupation of black holes in dark matter haloes, we find that  the number of gravitational wave sources detectable by SKA PTA is about $10^5-10^6$  at $ z \gtrsim 6$.  With their orbital velocities approaching $0.2 c$, a host of techniques including relativistic Doppler boosting leading to periodic variability in the optical light curves can be used to identify the electromagnetic counterparts of detected gravitational wave events. 

Using currently available constraints on the black hole  mass - circular velocity relation at high redshifts, combined with the latest observations of the quasar luminosity function at $z > 6$, we predicted the expected number of active quasar counterparts to SKA PTA detections, finding that the electromagnetic counterparts of the most massive black holes at $z \sim 6$ should be {\it uniquely identifiable} in the SKA localization ellipse. In so doing, we also developed data-driven estimates for the number of active quasar counterparts to the PTA events, as a function of the quasar's Eddington luminosity and active lifetime -- two parameters which are directly related to the black hole growth mechanism \citep{farina2022}. This allowed us to derive the expected numbers of quasar counterparts of known black holes at $z \gtrsim 6$ if the system were to be in a binary. Our results imply that identifying the counterparts of PTA events with the SKA will be able to place robust constraints on the seeding and growth mechanisms of high-$z$ SMBHs.  
  
\section*{Acknowledgements}  
 HP's research is supported by the Swiss National Science Foundation via Ambizione Grant PZ00P2\_179934. The work of AL is supported in part by the Black Hole Initiative, which is funded by grants from the John Templeton Foundation and the Gordon and Betty Moore Foundation. { We thank the referee for a helpful report.}

\def\aj{AJ}                   
\def\araa{ARA\&A}             
\def\apj{ApJ}                 
\def\apjl{ApJ}                
\def\apjs{ApJS}               
\def\ao{Appl.Optics}          
\def\apss{Ap\&SS}             
\def\aap{A\&A}                
\def\aapr{A\&A~Rev.}          
\def\aaps{A\&AS}              
\def\azh{AZh}                 
\def\baas{BAAS}
\def\jcap{JCAP}
\def\jrasc{JRASC}             
\def\memras{MmRAS}
\def\na{New Astronomy}
\def\nat{Nature}
\def\mnras{MNRAS}             
\def\pra{Phys.Rev.A}          
\def\prb{Phys.Rev.B}          
\def\prc{Phys.Rev.C}          
\def\prd{Phys.Rev.D}          
\def\prl{Phys.Rev.Lett}       
\def\pasp{PASP}               
\def\pasj{PASJ}
\def\physrep{Phys. Repts.}
\def\qjras{QJRAS}             
\def\skytel{S\&T}             
\def\solphys{Solar~Phys.}     
\def\sovast{Soviet~Ast.}      
\def\ssr{Space~Sci.Rev.}      
\def\zap{ZAp}                 
\let\astap=\aap
\let\apjlett=\apjl
\let\apjsupp=\apjs

\small{
\bibliographystyle{aa}
\bibliography{mybib2, mybib}
}

\end{document}